\documentclass[12pt]{article}

\usepackage{slashed}

\begin{document}

\title{ Higher order effective interactions and effective bosonized model for  2-N particle
 states \thanks{Contribution for the 21st International Conference on Few-Body Problems in Physics, Chicago, May 2015}}

\author{Fabio L. Braghin \thanks{braghin@ufg.br}
\\
Instituto de F\'\i sica, Federal University of Goias,
\\
Av. Esperan\c ca, s/n,
 74690-900, Goi\^ania, GO, Brazil
          }

\maketitle

\abstract{
 In this work, 
an effective  fermion model with particular  higher order 
interactions 
given by:
$I_{II} = \sum_n^N  g_{2^n} (\bar{\psi}_a \psi_a)^{2^n}$, for finite $N$,
is investigated by means of the auxiliary field  method
by taking into account 2$n$-particle states
as  
proposed in Ref. \cite{FLB-2015}.
The bosonized model was  found to exhibit an approximate symmetry 
 when expanded
for weak field fluctuations  around the 
mean field solutions.
In the present work,  the role of the mean field solutions for the corresponding
 auxiliary fields is investigated.
With the integration of fermions, 
the resulting determinant is expanded in a
 polynomial boson model in the weak field approximation
 and the    approximate symmetry found for this series 
does not appear
 if the boson  mean fields are considered to be zero.
}

\section{Introduction}
\label{intro}

The auxiliary field method \cite{kashiwa1,GN,kleinert2011,negele-orland}
is a powerful non perturbative method
which reduces the non linearities of a theory at the expenses of introducing
interacting
auxiliary fields (a.f.).
It has been successefully  applied in different systems from
quantum field, strong interacting  and nuclear systems
 models to  many body atomic and electronic systems
 \cite{negele-orland,klevansky,lp4,kleinert,braghin-navarra}.
 N-body effective interactions  can contribute in few and many body systems
and it  is one extremely difficult problem 
to take them into account.
Lately it has been shown that 
these higher order interactions constitute  important part of the  Effective Field Theory
 approach (EFT) \cite{EFT,EFT2}.
Some  new few body states and interactions have been observed in several systems in the last years 
such as in meson spectroscopy with the appearance of tetraquark states
and in cold atoms.
A  particularly interesting observed feature 
is an apparent approximate degeneracy of 2-particle and 4-particle states
 in light scalar mesons (masses below/around  1 GeV) \cite{scalars}
and also in 2 and 3 or 4 particle states in cold atoms by means of 
their binding energies \cite{cold-atoms}. 
Even though this last systems are non relativistic one might
search for  general feature(s)  of few body states built with 
few body interactions.
In the present work a complementary calculation to Ref. \cite{FLB-2015}
is performed based in a dynamical fermion effective model.
In the present work,  the particular series addressed in 
\cite{FLB-2015} given by
$I_{II} = \sum_n^N  g_{2^n} (\bar{\psi}_a \psi_a)^{2^n}$,
for finite $N$,   is analysed within the 
auxiliary field   method 
 if there are no
mean field solutions.
The logics of the work  is the following.
A dynamical model with higher order interactions is considered
and it is bosonized by introducing a set of N auxiliary 
fields.
Although from EFT  power counting 
or renormalization group arguments
the more relevant terms of a series of interactions
are not those present in the series $I_{II}$, the idea of the present 
development is the following:
to investigate the role of separate  terms or classes
 of higher order effective  interactions 
by switching on and off the others to understand 
their roles in the bosonized resulting effective model,
which describes sets of 2$n-$ particles composite states.
The minimum number of a.f. is introduced  such as to cancel out
all the interacting terms by means of minimum number of shifts,
which are considered to be the simplest ones.
Although this approach is strictly valid for weak 
fluctuations of the a.f. with respect to a coupling constant scale,
it was shown in Ref. \cite{FLB-2015} that it yields the correct expressions
if this weak field approximation is lifted.
By integrating out the original fermion fields a bosonized model for 
a collection of N-particle states is obtained which can be expanded
in polynomial boson effective interactions.
The saddle point equations of the bosonized model are considered to have
only trivial solutions without the formation of mean fields differently from
Ref. \cite{FLB-2015}.
The  resulting polynomial effective model is found 
and compared to the case in which there is non trivial
a.f.  mean field (condensates).
It will be shown that the new approximate symmetry between
N-fermion states appear only if the condensates  (mean fields)
are formed, i.e. if mean fields are taken into account.

\section{Auxiliary fields and bosonization}
\label{sec-1}

Consider the following generating funcional with a polynomial effective model:
\begin{eqnarray} \label{Z-L-original-2}
Z =  \int {\cal D} [\bar{\psi}_a, \psi_a]
e^{  \; i \int_x {\cal L}_{II} [\psi_a, \bar{\psi}_a]} ,
\end{eqnarray}
where the following  series of   bilinears   will be considered:
\begin{eqnarray} \label{series-II}
{\cal L}_{II} =
\bar{\psi}_a (x) \left( i \slashed{\partial} - m_a \right) \psi_a (x)
+ \sum_{n=1}^{N}  g_{2^n} \left( \bar{\psi}_a \Gamma \psi_a \right)^{2^n} ,
\end{eqnarray}
 where  $N$   even  and $\Gamma = I$
will be addressed; 
where $g_{2^n}$ are the effective coupling constants with 
dimension: $[g^{2^n}] = M^{d- (d-1)2^n}$, where $M$ has dimension of mass,
$m_a$ are the masses for each of the fermion species
and the index 
$a=1...N_r$ stands for the fermion components.
The auxiliary fields, with dimension $[M^{2}]$, are introduce by means of
 $N/2$  unity integrals that are given by:
\begin{eqnarray} \label{gaussian}
N'  \int \; {\cal D} [\varphi_n]
 e^{- i \int_x \frac{1}{2} \sum_n^{N} \frac{1}{d_n} \varphi^2_n (x)} 
 = 1 ,
\end{eqnarray}
The simplest necessary shifts of the auxiliary fields that cancel out the interactions  
can be written as:
\begin{eqnarray} \label{shifts-4n-1}
\varphi_n^2 \to (\varphi_n - \beta_n (\bar{\psi}_a \psi_a)^{2^{n-1}} 
)^2,
\end{eqnarray}
where $\beta_n$ are dimensionful parameters that are determined 
by imposing the corresponding cancelations of all polynomial  interactions.
There are other possible shifts in the auxiliary fields, however 
these are the simplest ones with the minimal number of
auxiliary fields that cancel out all the polynomial interactions
without introducing further previously non existing interations.
This reduces eventual ambiguities in the procedure.

For an arbitrary $n$, these conditions can be written in the following form:
\begin{eqnarray} \label{betas-g}
g_{2^n} &=&  - \frac{\beta_{n+1} }{ d_{n+1}} \varphi_{n+1}
+
\frac{\beta_n^2}{2 d_n} 
,
 \;\;\;\;  \mbox{for n $<$ N} ,
\nonumber
\\
g_{2^n} &=& 
\frac{\beta_n^2}{2 d_n} , \;\;\;\;\;\;\;\;\;\;\;\;\;\;\;\;\;\;\;\;\;\;\;\;\;\;\;\;\;
\mbox{for n $=$  N}.
\end{eqnarray}
From these expressions the parameters $\beta_n$ might become functions 
of the fields $\varphi_{n+1}$ without changing the unity Jacobian.
This minimal procedure   is  valid when all the fermion coupling constants, 
except $g_{2^N}$, are quite strong 
and  
 the subset of  $\varphi_{N-1}$ fields only assume positive values
and/or
 these auxiliary fields are weak with respect to 
the mean field which are weaker than the (normalized) fermion coupling constants.
This means that higher order auxiliary fields, which are introduced to 
cancel out progressively more irrelevant fermion interactions,
are  progressively weaker,
 i.e.  
$|\varphi_m \beta_m | <  g_{2^{m-1}}$ for positive coupling constants.

The resulting effective  action for fermions interacting with the auxiliary fields
 is given by:
\begin{eqnarray} \label{Seff-n}
S_{eff} &=&  \int d^4 x \;  
\left[
\bar{\psi}_a\left( i \slashed{\partial} - m_a + \frac{\beta_1}{d_1} \varphi_1 
\right) \psi_a - \sum_{n=1}^N \frac{1}{2 d_n} \varphi^2_n \right].
\end{eqnarray}
The saddle point equations for these auxiliary fields 
 provide   relations between the 
ground state average of the auxiliary fields 
the progressively large power of bilinears, i.e.
$\varphi_n \sim ( \bar{\psi}_a \psi_a )^n$, with an implicit
summation on the index $a$.
Therefore the auxiliary fields represent  corresponding 
scalar composite fermion  states, basically as a sum of states.
By integrating out fermions and  by using that 
$\det A = \exp\left( Tr \ln A \right) $
it yields:
\begin{eqnarray} \label{Seff-n}
S_{eff} &=& -  i Tr \log \left( i \slashed{\partial} - m_a + \frac{\beta_1}{d_1} \varphi_1 
\right) - \sum_{n=1}^N \int_x \frac{1}{2 d_n} \varphi^2_n,
\end{eqnarray}
where $Tr$  is the traces taken over all the internal indices and integration over space-time.
According to expression (\ref{betas-g}), there is a recursive dependence on all the 
a.f.  in the determinant because:
$\beta_1 [\varphi_n]$.
Only the trivial zero solutions for the gap equation 
will be considered.

\section{ Expanded effective boson model}

The first terms expansion corresponds to:
\begin{eqnarray} \label{deriv-exp}
 S_{eff} &\simeq&
S_{eff,(0)} [\varphi_i^{(0)}] 
+  \sum_i^N
\sum_j^N
 \frac{1}{n_i ! n_j !}
 \int_{x_1, x_2} 
  \left.
\frac{\delta^2 S_{eff}}{\delta \varphi_i (x_1) \delta  \varphi_j(x_2)} 
 \right|_{{\varphi_i}=\varphi_i^{(0)} }
\varphi_i (x_1) \varphi_j (x_2) 
+
h.o.
\end{eqnarray}
where $\int_{x_1, x_2} = \int d x_1 \int d x_2$, 
$h.o.$ 
stands for 
(even) higher order  derivatives, $n_i, n_j$ in the second derivative  term
and $n_i,n_j,n_k$  in the third order are such that  $n_i+n_j=2$ and 
$n_i + n_j + n_k = 3$ respectively. The first derivative term
is set to zero due to the stability condition. 

By assembling the interaction terms,
 it yields for the first three auxiliary fields:
\begin{eqnarray} \label{eff-I}
{\cal V}^{eff}_I &=& 
\sum_{j=2} \frac{ c_j }{2}  
\varphi_1^j
+ 
\sum_{j,k=1} t_{j,k}
\varphi_1^j   \varphi_{2}^k
+ 
\sum_{j,k,i=1} 
t_{j,k,i} \varphi_{1}^j \varphi_{2}^k \varphi_3^i
\end{eqnarray}
where $c_j$   are the resulting self interaction coupling 
constants for $\varphi_1$,
$t_{j,k}$ for $j+k \geq 2$ and
$t_{j,k,i}$ for $j+k+i \geq 3$.
It is seen that only the first 
a.f. develops a mass term,
and higher order fields only 
couple to successively lower order a.f.,
i.e.  $\varphi_n$ only develops couplings
to all the a.f. of the type $(\prod_m^{n-1} \varphi_m  \varphi_n)$.
This is a result of 
the zero condensates case.
For the sake of comparison the following expression 
obtained in Ref. \cite{FLB-2015} is reproduced 
here for the case of non zero condensates
in the case the fields are  suitably redefined to incorporate:
\begin{eqnarray} \label{cond-II}
{\cal V}^{eff}_I &=& 
\frac{1}{2}  \left[  c_2 
\chi_1^2
+  \left( c_2 + 
c_{2,2}
\right)
\chi_2^2
 + 
 \left( c_2 
 + c_{2,2} 
+ c_{2,3} 
\right)
\chi_3^2  
+ 
 \left( c_2 + c_{2,2} + c_{2,3} + c_{2,4}
\right)
\chi_4^2  
+ ... \right]
\nonumber
\\
&& + \sum_{n\geq 3} 
\left[  c_{ n}  \chi_1^n +  ( c_n + c_{n,2} )\chi_2^n + 
 ( c_n + c_{n,2} + c_{n,3} )\chi_3^n  + ... \right]
+ \sum_{i,j,k}
 t_{i,j,k}
\chi_1^i   \chi_2^j \chi_3^k ,
\nonumber
\\
&& t_{i,j,k} \;\;\; 
\mbox{for $ i+ j + k  \geq 2$, at least two indices non zero}
,
\end{eqnarray}
where $c_n$ and $c_{n,m}$ are the resulting self interaction coupling 
constants and contributions for masses, and also the 
couplings $t_{i,j,k}$ are those couplings between at least two different components,
being that at least two indices are non zero, i.e. $i,j\neq 0$ or $i,k \neq 0$ and so on.
From this expression the limit of progressively relatively large 
condensates yielded an effective potential with an approximate
symmetry.
By comparing
expressions (\ref{eff-I}) and (\ref{cond-II}) it is seen
there is a complete absence of the 
 explicit effective and approximate
symmetry that was found in the case of 
non zero (relatively large) higher order condensates \cite{FLB-2015}.

\section{ Conclusions}

In this work an effective bosonized model for $n$-fermion scalar states was
built from a dynamical effective fermion model with very particular 
higher order interactions
by means of the auxiliary field method
in the case of zero mean fields.
By integrating out fermions the resulting 
determinant was expanded in polynomial boson effective interactions.
The approximate symmetry found in Ref. \cite{FLB-2015} in the case 
of relatively large higher order condensates (mean fields) was found to 
be completely  absent in 
the case of zero scalar fields condensates.
The resulting effective potential (\ref{eff-I}) exhibits 
a preferential role  for the lowest order scalar a.f. ($\varphi_1$)
which appears  in all interaction terms.

\centerline{\bf Acknowledgement}
The author thanks short discussion with 
P. Bedaque,
and financial support from FAPEG-Goias and CAPES-MCT and CNPq, Brazil.

\end{document}